\def\comment#1{}
\def\old#1{}
\def\mn#1{*\marginpar[*#1]{*#1}}
\def\mn#1{}
\documentstyle[aps,epsfig,prl]{revtex}

 \def\lfrac#1#2{{{{#1}/{#2}}}}
\begin{document}

\title{ Thermodynamics of Crossover from Weak- to
Strong-Coupling Superconductivity}

\author{Egor Babaev\footnote {
email: egor@teorfys.uu.se \ http://www.teorfys.uu.se/people/egor/   \
Tel: +46-18-391902, Fax +46-18-533180
}}

\address{
Institute for Theoretical Physics, Uppsala University
Box 803, S-75108 Uppsala, Sweden }


\maketitle
\begin{abstract}
In this paper we study an
evolution of low-temperature thermodynamical quantities
for
an electron gas with
a $ \delta $-function attraction as the system
crosses over  from weak-coupling (BCS-type) to
strong-coupling (Bose-type) superconductivity
in three and two dimensions.
 \end{abstract}
\section{Introduction}
Recently there was  remarkable progress in 
 the study of the crossover from
weak-coupling (BCS) superconductivity to Bose-Einstein condensation
(BEC)
of tightly bound fermion pairs. Although this crossover
was first addressed
a long time ago \cite{Ea}-\cite{Le}, the recent
interest in this subject \cite{R8}-\cite{Micnas} was initiated
by suggestions of its relevance to
High-Temperature Superconductors \cite{u5}-\cite{ranrev}.
This  crossover is a remarkable example of smooth evolution between
two very different physical systems. A few of its
features are: in the BCS superconductors there is one characteristic
temperature $T_c$ where superconductivity disappears
due to thermal decomposition of Cooper pairs. In contrast, the
``Bose-type"  superconductor possesses two characteristic temperatures,
namely a critical temperature which in the extreme Bose regime
does not depend on coupling strength
and a substantially higher temperature
of thermal pair decomposition. Also the BCS condensate is characterized
by a coherence length which coincides with the size of a Cooper pair
whereas in the Bose limit the system possesses two characteristic length scales,
namely a coherence length which increases with increasing coupling strength
and a Cooper pair size which obviously decreases
with increasing coupling strength.

In fact, this crossover is related to many other
phenomena in various branches of physics. In particular
it was generalized to the Chiral Gross-Neveu (GN) model \cite{gn1,gn2}.
This model  is  considered as a toy model for QCD.
There are also numerous ongoing  
discussions of the possibility of the appearance of  a pseudogap phase
in the Nambu--Jona-Lasinio (NJL) model \cite{njl1,njl2,njl3}\footnote{
The Gross-Neveu and Nambu--Jona-Lasinio
models are models of N-component fermions fields
that serve for description of dynamic chiral symmetry breakdown
in particle physics.
Originally the NJL model was proposed in analogy to BCS
theory. The NJL model
may be treated by mean-field methods in the regime
of large number of colors $N_c$ with $1/N_c$ serving
as  a small parameter of the problem. However
in a real world $N_c=3$ thus as it is not infinite,
 the system may possess [e.g. at finite temperature]
a fluctuation-driven  restoration of the chiral symmetry
while preserving nonzero constituent quark mass. This would be
a phenomenon similar to the symmetry restoration
without pairbreaking
in strong coupling or low carrier density superconductors.
However in the case of  the NJL model there are complications
that do not allow this problem to be addressed rigorously by
perturbative methods \cite{njl1,njl2,njl3}.
Thus BCS-BEC crossover
in superconductors may in some sense serve as  a valuable
toy model for QCD. }.

In contrast to the particle physics, in the
theory of superconductivity the BCS-BEC crossover
has been studied in great detail.
However, the main subject of the
study was  the evolution of critical
temperature and non-Fermi--liquid properties above $T_c$.
Intuitive clarity of the physical picture of this crossover
is obscured by the
absence of simple analytical and asymtotic results 
for many other physical quantities (e.g.
for the behavior of thermodynamic functions).
The low-temperature asymptotics of thermodynamic
functions of a BCS superconductor
are well known and can be found in many textbooks.
However to the best of the author's knowledge there are
no analogous expressions published for the
thermodynamics of BCS-BEC crossover.
This paper is an attempt to fill in this gap.

In the model to be investigated in this paper,
the crossover
from BCS-type to Bose-type
superconductivity
takes place either by
varying the coupling
strength, or by
decreasing
the carrier density.
Asymtotic calculations in the BCS and BEC limits 
will be performed
using the crossover parameter $x_0$ that we adopt from
\cite{M}. This parameter
is directly related to the ratio of the chemical potential and 
gap function at zero temperature. It
is a monotonic function of coupling strength and
carrier density
[to be seen in Figs.~\ref{f1}
and \ref{f2} in two and three dimensions, respectively].
It is also a
direct measure for the
scattering length $a_s$ of the electron-electron
interaction, at zero temperature
 in three dimensions [to be seen in Eq.~(\ref{1.8})], and
of the binding energy of the
electron-electron pairs in two dimensions [to be seen in
Eq.~(\ref{@binding})].

The plan of the paper is following:
we start by a brief outlining of the
results of Refs.~ \cite{M},
 \cite{2Danalit} for the gap $ \Delta (0)$
and chemical potential at zero temperature in the entire
crossover region.
These are subsequently extended
by equations for the low-temperature ($T/\Delta(T) <<1$)
behavior of  the gap
functions, chemical potential 
and  of thermodynamic functions in two and three dimensions
in the BCS and BEC limits up to the boundaries
of the region $-1<x_0<1$ .

In appendix we also calculate the
evolution of thermodynamic quantities
near $T_c$
in weak and moderately strong coupling superconductors where
mean-field methods are reliable.
\section[The Model ]{The Model}
\comment{
\section[ Thermodynamics of
the Crossover from BCS to Strong-Coupling Superconductivity  Near Zero
Temperature]
{Thermodynamics of
the Crossover from BCS to Strong-Coupling Superconductivity  Near Zero
Temperature}}

The Hamiltonian of our model is the typical BCS Hamiltonian
in $D$ dimensions ($\hbar=1$)
\begin{eqnarray}
H &=& \sum_\sigma \int \! d^D x
        \, \psi_\sigma^{\dag} ({\bf x})
        \left(-{{\bf \nabla}^2 \over 2m} -\mu\right)
        \psi_\sigma({\bf x})
         + g \int\!d^D x\,
        \psi_\uparrow^{\dag}({\bf x}) \psi_\downarrow^{\dag}({\bf x})
        \psi_\downarrow^{\phantom{\dag}}({\bf x})
        \psi_\uparrow^{\phantom{\dag}}({\bf x})
        \label{1.0},
\end{eqnarray}
where $\psi_\sigma({\bf x})$ is the Fermi field operator,
$\sigma=\uparrow,\downarrow$
denotes the spin components,
$m$ is the effective fermionic mass, and
$g < 0 $  the strength of an attractive potential
$ g  \delta ({\bf x} - {\bf x}')$.

The mean-field equations for the
gap parameter $\Delta$ and the chemical potential $\mu$
are obtained  in the standard way from the equations:
\begin{eqnarray}
-{1\over g} &=& \frac{1}{V} \sum_{\bf k} {1\over 2 E_{\bf k}}
\tanh{E_{\bf k} \over 2T} ,\label{1.1}\\
  n &=& {1\over V} \sum_{\bf k} \left(1-{\xi_{\bf k}
 \over E_{\bf k}} \tanh{E_{\bf k} \over 2T}\right),
\label{1.2}
\end{eqnarray}
where the sum runs over all
wave vectors
$\bf k$,
 $N$ is the total number
of fermions,
 $V $  the
volume of the system,
 and
 $E_{\bf k}=\sqrt{\xi_{\bf k}^2 + \Delta^2}$
with
$\xi_{\bf k} = {\bf k}^2 / 2 m - \mu$
%
 are the energies of single-particle excitations.

Changing the  sum over {\bf k} to  an integral
over $\xi$
and over the directions of ${\bf k}$, on which the integrand does not depend,
 we arrive
in three dimensions at the gap equation:
\begin{equation}
{1 \over g} = \kappa_{3} \int_{-\mu}^\infty d\xi {\sqrt{\xi+\mu}
\over 2 \sqrt{\xi^2+\Delta^2}} \tanh{\sqrt{\xi^2 +\Delta^2} \over 2 T}
\label{1.3.1},
\end{equation}
where the constant $\kappa_{3}= m^{3/2} / \sqrt{2} \pi^2$ has dimension
energy$^{-3/2}$/volume.
In two-dimensions, the  density of
states is constant, and the gap equation becomes
\begin{equation}
{1 \over g} = \kappa_{2} \int_{-\mu}^\infty d\xi  { 1
\over 2 \sqrt{\xi^2+\Delta^2}} \tanh{\sqrt{\xi^2 +\Delta^2} \over 2 T}
\label{1.3.2},
\end{equation}
with a constant $ \kappa_{2}= m/2\pi$  of dimension energy$^{-1}$/two-volume.
In two dimensions, the particle number in Eq.~(\ref{1.2}) can
be integrated
with the result:
\begin{equation}
n=\frac{m}{2 \pi} \left\{
\sqrt{\mu^{2} + \Delta^{2}} + \mu +
2 T \log{\left[
1 + \exp{\left(-\frac{\sqrt{\mu^{2} + \Delta^{2}}}{T}\right)}
\right]}\right\}.
\label{numb}
\end{equation}

The $\delta$-function potential produces an
artificial divergence and requires
 regularization. A BCS superconductor possesses
 a natural cutoff supplied by the Debye frequency $ \omega _D$.
For the crossover problem
to be treated here
this is no longer a useful quantity, since in the strong-coupling
limit all
 fermions
participate in the interaction, not  only those
in a thin shell of width $ \omega _D$ around the Fermi surface.
To be applicable in this regime,
 we  renormalize
the gap equation in three dimensions with the help of the
experimentally observable
$s$-wave scattering length $a_s$,
for which the  low-energy limit of the
two-body scattering process gives an equally divergent
expression:
\begin{equation}
        {m \over 4 \pi a_s}
=
{1\over g}
        + {1\over V}
\sum_{\bf k}
{m \over {\bf k}^2} .
 \label{1.4}
\end{equation}
Eliminating $g$ from (\ref{1.4}) and  (\ref{1.1})
we obtain a renormalized gap equation
\begin{equation}
        -{m \over 4 \pi a_s} = {1\over V} \sum_{\bf k}
        \left[{1\over 2 E_{\bf k}} \tanh{E_{\bf k} \over 2T}
 - {m \over {\bf k}^2} \right],
        \label{1.5}
\end{equation}
 in which $1/k_Fa_s$ plays the role
of a dimensionless coupling constant which monotonically increases
 from $-\infty$ to $\infty$ as the bare
coupling constant $g$ runs from small
(BCS limit) to  large values
(BE limit).
This equation is to be solved
simultaneously with (\ref{1.2}).
\comment{
These mean-field equations were
first analyzed
at a fixed carrier density in Refs.~\cite{R8} and \cite{R-rev}.
Here we shall first
reproduce the obtained
estimates for $T^*$ and $\mu$.}
In the BCS limit, the chemical potential $\mu $
does not differ much
from the Fermi energy
 $\epsilon_F$, whereas with increasing interaction strength,
 the distribution function $n_{\bf k}$
broadens and $\mu $ decreases, and
in the BE limit, on the other hand  we have tightly bound
pairs and nondegenerate fermions with a large negative chemical
potential $|\mu|\gg T$.

\begin{figure}[tbh]
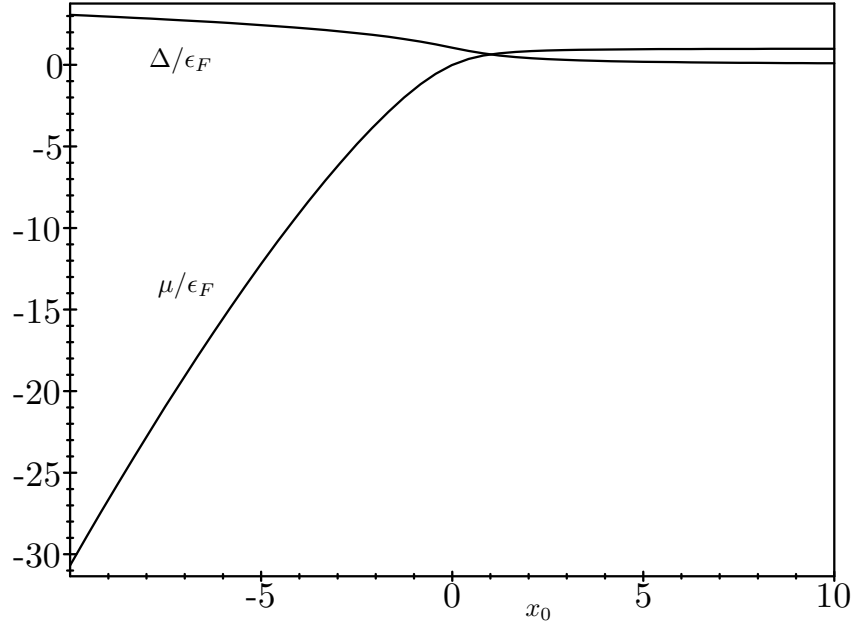

\input 3D0.tps
\caption{
Gap function $\Delta$
and chemical
potential $\mu$ at zero temperature
as functions of $x_0 = \mu / \Delta$ in three dimensions.
}
\label{f1}\end{figure}

In three dimensions at $T=0$, equations (\ref{1.5}), (\ref{1.2})
were solved analytically
in entire crossover region in \cite{M} to obtain
$\Delta$ and $\mu$ as functions of
crossover parameter $1/k_Fa_s$. The results are
\begin{eqnarray}
        {\Delta \over \epsilon_F}
        &=&
        {1\over [x_0 I_1(x_0) + I_2(x_0)]^{2/3}}   ,
        \label{1.6}
        \\
        {\mu \over \epsilon_F}
        &=&
        {\mu \over \Delta} {\Delta \over \epsilon_F}
        = {x_0 \over [x_0 I_1(x_0) + I_2(x_0)]^{2/3}},
        \label{1.7}
        \\
        {1\over k_F a_s}
        &=&
        - {4\over \pi} {x_0 I_2(x_0) - I_1(x_0)
        \over [x_0 I_1(x_0) + I_2(x_0)]^{1/3}}        ,
        \label{1.8}
\end{eqnarray}
with the functions
\begin{eqnarray}
        I_1(x_0)
        =
        \int_0^\infty \!\!\!\! dx
        {x^2 \over (x^4 - 2 x_0 x^2 + x_0^2 +1)^{3/2} }
                =
        (1+x_0^2)^{1/4}
        E({\scriptstyle \pi\over \scriptstyle 2},\kappa) - {1\over 4
x_1^2(1+x_0^2)^{1/4}}
        F({\scriptstyle \pi\over \scriptstyle 2}, \kappa), \nonumber
\end{eqnarray}
\begin{eqnarray}
        I_2(x_0)
        =
        {1\over 2} \int_0^\infty\!\!\!\! \,dx
        {1\over (x^4 -2x_0 x^2 + x_0^2 +1)^{1/2}}
        \nonumber
        =
        {1\over 2 (1+x_0^2)^{1/4}}
        F({\scriptstyle\pi\over \scriptstyle 2}, \kappa)  ,
\nonumber
\end{eqnarray}
\begin{eqnarray}
        \kappa^2
        =
        {x_1^2 \over (1+x_0^2)^{1/2}}                      ,
        x^2
        =
       {k^2 \over 2m}{1 \over \Delta} \ , \ \  \ \ x_0= {\mu \over \Delta},
   \ \ \ \ x_1^2 = \frac{\sqrt{1+x_0^2}+x_0}{2} ,
\nonumber
\end{eqnarray}
where  $E({\pi\over 2},\kappa)$ and $F({\pi\over 2},\kappa)$ are
the usual elliptic integrals. The quantities
(\ref{1.6}) and (\ref{1.7}) are plotted
as functions of the crossover parameter $x_0 = \mu / \Delta$
in Fig.~\ref{f1}.

\begin{figure}[tbh]
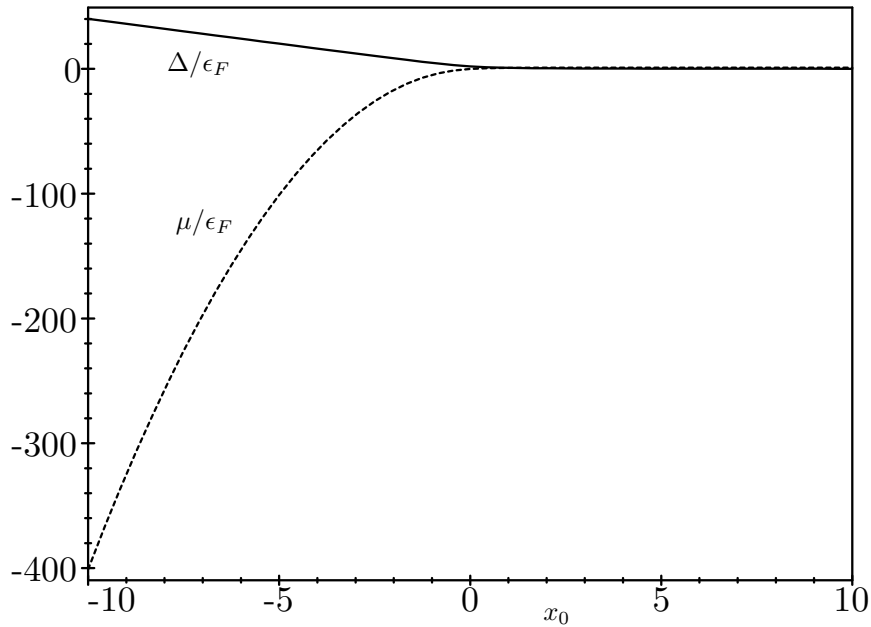

\input 2D0.tps
\caption{
Gap function $\Delta$
and chemical
potential $\mu$ at zero temperature
as functions of $x_0$
in two dimensions.
}
\label{f2}\end{figure}

In two dimensions, a nonzero bound state energy $\epsilon_0$
exists for any coupling strength. The cutoff
can therefore be eliminated by
subtracting from the two-dimensional zero-temperature gap equation
\begin{equation}
- \frac{1}{g} = \frac{1}{2V} \sum_{\bf k} \frac{1}{\sqrt{\xi_{\bf k}^2 +
\Delta^2}}=
\frac{m}{4 \pi} \int_{- x_0}^{\infty} d z \frac{1}{\sqrt{1+ z^2}},
\end{equation}
where $ z=k^2/2m\Delta-x_0$,
the bound-state equation
\begin{equation}
-\frac{1}{g}=\frac{1}{V}\sum_{\bf k} \frac{1}{ {\bf k}^2/m + \epsilon_0}=
\frac{m}{2\pi}\int_{-x_0}^\infty d z \frac{1}{ 2z+\epsilon_0/\Delta+2x_0}.
\label{@binding}\end{equation}
After performing the elementary integrals,
one finds :
${\epsilon_0}/{\Delta}=\sqrt{1+x_0^2} - x_0.$
{}From Eq.~(\ref{numb}) we see that at zero temperature,
gap and chemical potential
are related to $x_0$ by \cite{2Danalit,M}
\begin{eqnarray}
   {\Delta \over \epsilon_F }
   &=&
   {2 \over x_0 +\sqrt{1+x_0^2}},
   \label{1.13}
\end{eqnarray}
\begin{eqnarray}
   {\mu \over \epsilon_F}
   &=&
   {2 x_0 \over x_0 +\sqrt{1+x_0^2}}.
   \label{1.14}
\end{eqnarray}
The two relations are plotted in Fig.~\ref{f2}.
From (\ref{1.13})
we find
the dependence of the ratio $ \lfrac{\epsilon_0}{\epsilon_F} $
on the crossover parameter $x_0$:
\begin{equation}
\frac{\epsilon_0}{\epsilon_F}=
2 \frac{\sqrt{1+x_0^2} -x_0}{\sqrt{1+x_0^2} +x_0 }
\label{q1}
\end{equation}
\section[ The Crossover from BCS to Strong-Coupling Superconductivity  Near Zero
Temperature]
{The Crossover from BCS to Strong-Coupling Superconductivity  Near Zero
Temperature}
\subsection[The range of validity of mean-field
approximation]{The range of validity of mean-field approximation}
We have extended
the above relations to non-zero temperature
in the BCS and BEC limits. We use a mean-field
approximation 
which,
 in the range of its validity, allows  a
nice simple description of
the crossover.
It is known that in BCS limit the mean-field approximation 
is valid for the description of the entire superconductive region.
In fact the system crosses over from BCS to Bose
condensate in a very narrow region $-1<x_0<1$ (see e.g. \cite{pist})
so the BCS approximation is adequate for description of the entire
superconductive region in the range  $x_0 > 1$. 
This is in contrast to BEC limit where the superconductive transition 
is governed by the thermal excitation of collective modes rather than 
thermal decomposition of individual 
particles, the effect which is missed in the mean-filed 
approximation. So in the BEC limit we must be restricted in the 
following discussion to the temperatures much
lower than the temperature of superconductive transition:
 $0 \leq T << T_{Bose}$,
where $T_{Bose}=[n/2 \zeta(3/2)]^{2/3} \pi /m $  \ [see e.g. \cite{Noz,R8}].

A special care should be taken when a mean-field
approximation is used for the  description of a two-dimensional system.
It is  well-known,  that a straightforward calculations of the 
next-to-leading order
corrections gives that the gap should be
exactly zero at any finite temperature in 2+1 dimensions.
This is due to the fact that, according to standard 
dimensional reduction arguments, a 2+1 -dimensional system 
is effectively two dimensional at high temperature and 
in two dimensions there is no Goldstone boson as 
it is articulated in Coleman-Mermin-Wagner-Hohenberg theorem
\cite{col}. 
However as it was shown by Witten
\cite{W} (see also \cite{gn1,njl2,shar,sc,gn2}) a direct calculation of corrections
misses essential physics of a two-dimensional problem
and a mean-field results still have sense under certain 
conditions in 2D.
The situation  is following:
the fluctuations can be made arbitrarily weak by
decreasing temperature in 2+1 dimensions
(or e.g. increasing N
in 2D zero temperature  calculations in \cite{W,gn1,njl2})
and then employing a
``modulus-phase variables"  $\Delta=|\Delta| e^{i \theta}$
one can observe that 
the system possesses a very well-defined ``mexican hat"
effective potential that determines the gap { \it modulus}.
Due to phase fluctuations in the degenerate valley of the potential, the
average of the complex gap function is zero, however
there exist a gap locally (i.e. in some sense the system in
its low energy domain degenerates to a nonlinear sigma model).
In 2+1 dimensions,  as the temperature approaches
zero the thermal fluctuations in the degenerate valley
of the effective potential gradually vanish and at $T=0$
a local gap becomes a real gap. 
Existence
of the {\it local} gap modulus $\Delta$ does not contradict the
Coleman-Mermin-Wagner-Hohenberg
theorem since $\Delta$ is neutral under $U(1)$ transformations.
Thus at low temperature in 2+1  dimensions
there appears an ``almost" Goldstone boson
that becomes a real Goldstone boson at exactly zero temperature.
At low temperatures we can neglect contribution from
weak phase fluctuation at temperatures much lower than
the temperature of the Kosterlitz-Thouless transition, 
which in the weak coupling limit approximately coincides
with the temperature of  pair formation and in strong-coupling 
limit reaches a plateau value:
$T_{\rm KT} = {\epsilon_F}/{8} = {\pi} n/{8m}$,
where  $\epsilon_F=(\pi/m)n$ is 
the Fermi energy of free fermions
with the carrier density $n$ and mass $m$ \cite{Dr,Em,shar,sc}.

Let us now outline how finite temperature quantities will be derived.
We start with inspection of the gap $\Delta$ behavior at low temperature.
$\Delta(T)$ can be expressed with the help of the  analytic
zero-temperature results from \cite{M}. For example in three
dimension we can write an equation that relates the 
gap at zero temperature with the gap at finite temperature at a
fixed crossover parameter $x_0$ (which enters implicitly $\Delta(0)$):
\begin{equation}
\int_{-\mu}^\infty d\xi {\sqrt{\xi+\mu}
\over  \sqrt{\xi^2+\Delta(0)^2}} 
=
\int_{-\mu}^\infty d\xi {\sqrt{\xi+\mu}
\over  \sqrt{\xi^2+\Delta(T)^2}} \tanh{\sqrt{\xi^2 +\Delta(T)^2} \over 2 T}.
\label{new01}
\end{equation}
This equation should be solved self-consistently with eq. (\ref{1.2}).
We must observe that the above system of equations,
in contrast to zero-temperature calculations  has the temperature as
an extra parameter. Thus we need to specify in what model
we are studying temperature effects  with two most general options
being: (i) ``fixed chemical potential model" - i.e. when we 
do not fix the carrier density
but assume
the presence of a reservoir which provides us with a
temperature-independent chemical potential
$\mu=\mu( 1/k_Fa_s ; T=0 )$ or (ii) ``fixed carrier" density 
model - i.e. when $n$ is fixed and $\mu$ varies with temperature.
We choose the first option since such a fixed-$\mu$ model
is the most convenient
for deriving asymptotic
results for the finite-temperature behavior of the system.
However the choice of the model is not principal since 
from (\ref{1.2}) it is seen 
that in a ``fixed density" regime
the chemical potential in BCS and BEC limits
depends very weakly on temperature
in comparison with the dependence on the coupling strength
in the regimes where our formulas are valid.
Moreover the temperature dependence of 
the chemical potential in the entire crossover 
region
was calculated numerically in 2D in \cite{shar}
and it is small comparing to its
variation with coupling strength 
in the temperature range $0<T<T_{KT}$.
So the results are
similar in both cases of the presence of a particle
reservoir that provides temperature independent
chemical potential when carrier density 
varies or in the model of temperature independent
 carrier density.

\begin{figure}[tbh]
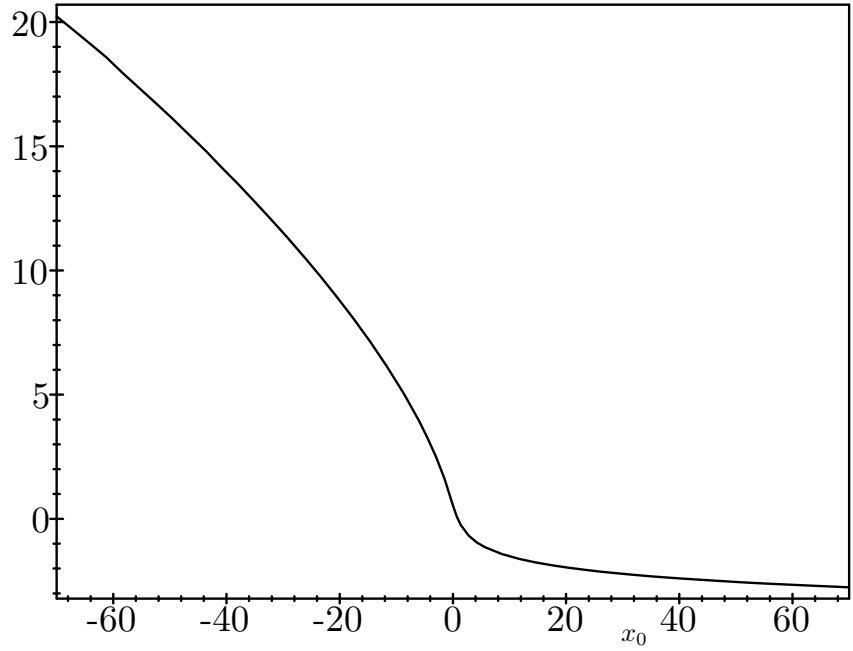

\input AS.tps  \\[-2cm]
\caption{
Dependence of $1/k_F a_s$ on the
crossover parameter $x_0$ in three dimensions
at zero temperature.
}
\label{f3a}\end{figure}
\begin{figure}[tbh]
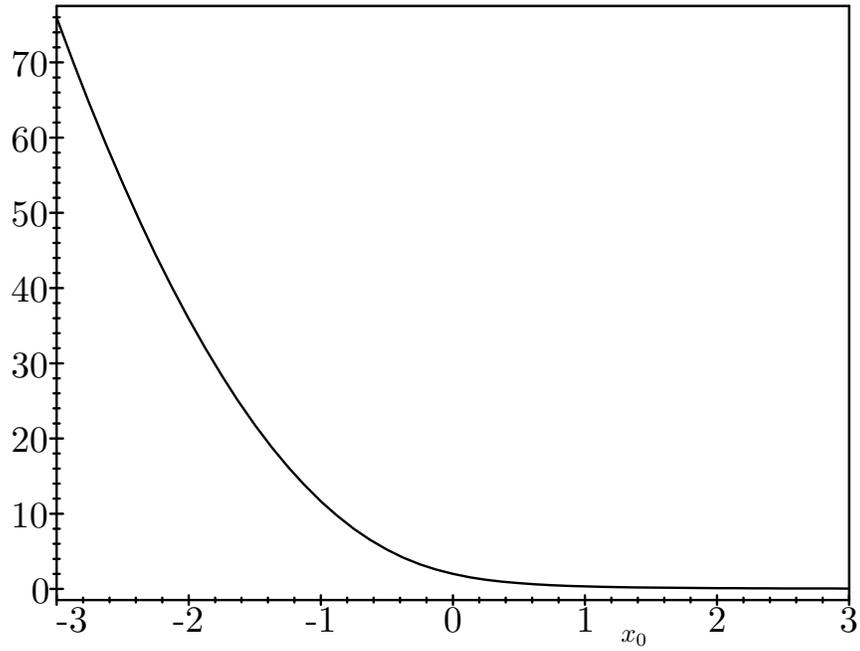

\input E0.tps  \\[-2cm]
\caption{
Dependence of $\epsilon_0/\epsilon_F$ on the
crossover parameter $x_0$  at zero temperature.
}
\label{f3b}\end{figure}

In our calculation we use $x_0$ as the most convenient
 crossover parameter,
since it depends
via
the simple relation $x_0=\mu /\Delta$
on the chemical potential which
can be measured  rather directly experimentally \cite{Rie}.
The parameter $x_0$
ranges from $-\infty$ in the strong-coupling (Bose-Einstein) limit
to $\infty$ in the weak-coupling (BCS) limit.
The relation between  $x_0$
 and the inverse reduced coupling strength
between the electrons $1/k_F a_s$
is plotted
for three-dimensional system
in Fig.~\ref{f3a}.
The corresponding relation
(\ref{q1}) in two dimensions
between $x_0$ and
the bound state energy $\epsilon_0$
of the electron pairs
is plotted on Fig.~\ref{f3b}.
\subsection{The gap functions}
Let us now turn to the region near zero temperature,
where the mean-field approximation gives 
an exact results for the
gap function.
 From (\ref{new01}) we extract the asymptotic behavior
in the three-dimensional case  at low $T$ having 
$ T/\Delta(T)$ as a small parameter.
For this purpose let us first rewrite (\ref{new01}) in three dimensions as:
\begin{equation}
\int_{-\mu}^\infty d\xi \left[{\sqrt{\xi+\mu}
\over  \sqrt{\xi^2+\Delta(0)^2}} 
-
{\sqrt{\xi+\mu}
\over  \sqrt{\xi^2+\Delta(T)^2}} \right]
= -
\int_{-\mu}^\infty d\xi {\sqrt{\xi+\mu}
\over  \sqrt{\xi^2+\Delta(T)^2}} 
\left[1-\tanh{\sqrt{\xi^2 +\Delta(T)^2} \over 2 T}\right].
\label{eneww01}
\end{equation}
In the region  $x_0>1$ one can assume density of states to be  roughly constant,
since the integrand of (\ref{1.3.1})
is  peaked
in the narrow region near $\xi=0$. 
Then the integral in the {\it r.h.s} of (\ref{eneww01})
can be  writen as
\begin{equation}
2\sqrt{\mu}\int_{-\mu}^\infty d\xi {1
\over  \sqrt{\xi^2+\Delta(T)^2}} 
\exp\left[-{\sqrt{\xi^2 +\Delta(T)^2} \over  T}\right].
\label{1111}
\end{equation}
This integral can be calculated e.g. by changing the integration varaible to
$y=\sqrt{(\xi/\Delta)^2+1}-1$ and expanding the factor before
exponent.
The result is expressed via the error function
and we arrive at the following expression
for the behavior of the gap
function near $T=0$ in the region $x_0>1$
%
%
\begin{eqnarray}
\Delta (T) = \Delta (0)  - \Delta (0) \sqrt{ {\pi \over 2} }
\sqrt{ { T \over \Delta (0) } }
\exp\left[- {  \Delta (0) \over T } \right]  \left[  1+ {\rm erf}
\left( \sqrt{\frac{\sqrt{x_0^2+1}-1}{T/ \Delta(0)}}
 \right) \right],
\label{3d0}
\end{eqnarray}
where ${\rm erf}(x)$ is the error function.
Since the density of states
is nearly constant in this limit,
the same equation  holds in two-dimensions---apart from  a modified gap
 $\Delta (0) $ given by (\ref{1.13}).

In the weak-coupling limit,
 $x_0=\lfrac{ \mu}{ \Delta (0) } $ tends to infinity,
and
the expression above approaches exponentially fast
the well-known BCS result:
\begin{eqnarray}
\Delta(T)
=\Delta(0) - [2 \pi \Delta(0) T] ^{1/2}\exp\left[-{ \Delta(0) \over T } \right]
\label{2dbcs}
\end{eqnarray}
For strong couplings with
 $ x_0 < - 1 $ the integration range, in contrast to the above case, does not
include the point $\xi=0$,
 so 
we write the {\it r.h.s} of (\ref{eneww01}) as:
\begin{eqnarray}
{2}{\sqrt{\Delta(0)}} \int_{-x_0}^{\infty}
\frac{\sqrt{\xi/\Delta(0)+x_0}}
{\sqrt{(\xi/\Delta(0))^2+(\Delta(T)/\Delta(0))^2}}
\exp\left[- \frac{\sqrt{(\xi/\Delta(0))^2
+(\Delta(T)/\Delta(0))^2}}{T/\Delta(0)}
\right]
d\left(\frac{\xi}{\Delta(0)}\right)
\end{eqnarray}
This integral can be calculated e.g. by changing the
integration variable to $v=\sqrt{(\xi/\Delta(0))^2+1} -\sqrt{x_0^2+1}$
and expanding the factor before exponent so that the 
above integral is reduced to:
\begin{eqnarray}
\frac{2\sqrt{\Delta(0)}}{|x_0|}
\exp\left[ -\frac{\sqrt{x_0^2+1}}{T/\Delta(0)}\right]
\int_0^{\infty} \exp\left[-\frac{ \Delta(0)}{T} v\right]\sqrt{v}dv
\end{eqnarray}
Thus, we find the behavior of the gap function at $x_0<-1$:
\begin{eqnarray}
\Delta(T) = \Delta(0) - { 8 \over \sqrt{\pi} }
\frac{1}{ \sqrt{-x_0}}
{T^{3/2} \over \sqrt{\Delta(0)}  }
 \exp
\left[ - \frac{\sqrt{\mu^2+\Delta^2(0)}}{T}
\right].
 \label{3dbe}
 \end{eqnarray}
 From Eq.~(\ref{3dbe}) we see that near $T=0$ the gap
$\Delta(T) $ tends
in the strong-coupling limit
 exponentially to $ \Delta(0) $,
forming plateau near $ T=0$.

In two dimensions
we arrive at similar result: an
exponentially growing plateau near $T=0$ in the
strong  coupling  limit:
\begin{eqnarray}
\Delta(T) = \Delta(0) - { \Delta(0) \over 2 }  E_1 \left(
{ \sqrt{\Delta(0)^2+\mu^2}  \over T } \right),
\label{2dbe}
\end{eqnarray}
where $E_1$ is the    exponential integral
$E_1 (z)= \int_z^{\infty} e^{-t}/t~ dt $.
\comment{
For very strong couplings, Eq.~(\ref{2dbe}) becomes:
\begin{eqnarray}
\Delta (T) = \Delta (0)- { \Delta (0) \over 2 }
{T \over \sqrt{\mu^2+\Delta^2(0)} }
{\rm exp} \left( - { \sqrt{\mu^2 + \Delta(0)^2} \over  T} \right)
 \label{extr}
 \end{eqnarray}}
The plateau formation
in the low temperature solution for the gap function in
a strong-coupling regime  (see Fig. 5) is an indication of the suppression of a pairbreaking
effects in low temperature domain. One can however
observe
that in contrast to BCS regime in the BEC regime the
suppression comes from a large negative chemical
potential in the exponent $|\mu| >> |\Delta|$.
\begin{figure}[tbh]
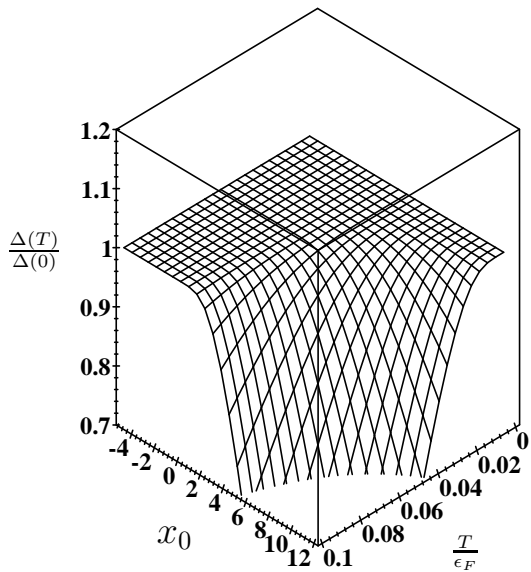

\input 3DN.tps  \\[-2cm]
\caption{``Plateau" formation of the 
gap function at low temperature in three dimensions.
In the BCS limit the thermal fluctuations start breaking fermion pairs
already at low temperatures which results in the 
absence of plateau of  $\Delta(T)/ \Delta(0)$. 
In the BEC limit in contrast we have a plateau
near zero temperature. Inclusion of Gaussian corrections in the 
BEC limit leads to depletion of the average order parameter 
due to depletion of macroscopic occupation of zero momentum level.
The order parameter in this limit becomes 
zero at $T_c=[n/2 \zeta(3/2)]^{2/3} \pi /m \approx 0.218 \epsilon_F$.
This region is not plotted on the above figure. }
\label{f3db}\end{figure}
\subsection{The thermodynamics}
Let us now calculate thermodynamical quantities
near $T=0$.
For the thermodynamic Gibbs potential
$ \Omega (T,\mu, V)$  we have
\begin{eqnarray}
 \Omega  = \sum_{\bf k} \left\{
{ \Delta ^2 \over 2 \sqrt{\xi^2_{\bf k} + \Delta^2 }  }
 \tanh  { \sqrt{\xi^2_{\bf k} + \Delta^2} \over 2T }  -
2 T \log \left[ 2 \cosh   { \sqrt{\xi^2 _{\bf k} + \Delta^2} \over 2 T }\right]
  +\xi_{\bf  k}
\right\}.
\label{3om}
\end{eqnarray}
Here and in the sequel in this section, $ \Delta(0) $ will be replaced by
$ \Delta $.
In three dimensions, Eq.~(\ref{3om}) turns into the
\begin{eqnarray}
\!\!\!\!\!\!\!\!\!{ \Omega  \over V }
 &\!\! = \!\!& \kappa_3 \int_{-\mu}^{\infty}  d\xi
 \sqrt{\xi + \mu}
\left[
{ \Delta ^2 \over 2 \sqrt{\xi^2 + \Delta^2 }  }
 \tanh
{  \sqrt{\xi^2 + \Delta^2} \over 2T }
 -
2 T \log \left( 2 \cosh
  {  \sqrt{\xi^2 + \Delta^2} \over 2 T } \right)
  +\xi
\right]\!,
\label{i3om}      \nonumber \\&&
\end{eqnarray}
In two dimensions, we obtain instead:
\begin{eqnarray}
{ \Omega  \over V }
=\kappa_2 \int_{-\mu}^{\infty} d\xi
\left[
{ \Delta ^2 \over 2 \sqrt{\xi^2 + \Delta^2 }  }
 \tanh
{  \sqrt{\xi^2 + \Delta^2} \over 2T }
-
2 T \log \left( 2 \cosh
  {  \sqrt{\xi^2 + \Delta^2} \over 2 T }
 \right)  +\xi
\right],
\label{i2om}
\end{eqnarray}
We regularize the thermodynamic potential $\Omega_s $
of the  condensate
subtracting $ \Omega_n= \Omega (\Delta = 0)$.
 At $T=0$ and  for weak couplings  this is found to depend
 on $\mu$ and $\Delta$ as follows:
\begin{eqnarray}
 \frac{ \Omega _s}{V} \equiv { \Omega - \Omega_n \over V} =
\kappa_3 \sqrt{\mu}  \left[ - { \Delta^2\over  4}
+{1 \over 2}  \mu | \mu | - { 1 \over 2} \mu \sqrt{\mu^2 +\Delta^2 }
  \right]
 \label{o3t0}
\end{eqnarray}
In the BCS limit ($x_0 \rightarrow \infty$) this reduces
to  the well-known  result
\begin{eqnarray}
{ \Omega_s  \over V} =
\kappa_3 \sqrt{\mu}  \left[ - { \Delta^2\over  2}
 \right].
\label{o3t0scs}
\end{eqnarray}
In two dimensions, we have a formula
valid for any strength of coupling or carrier density:
\begin{eqnarray}
{ \Omega _s \over V} =
\kappa_2  \left[ - { \Delta^2\over  4}
+{1 \over 2}  \mu | \mu | - { 1 \over 2} \mu \sqrt{\mu^2 +\Delta^2 }
 \right],
 \label{o2t0}
\end{eqnarray}
%
%
%

In the opposite limit of strong couplings,
 we find
in three dimensions:
\begin{eqnarray}
 {\Omega \over V}  =
- {\pi \over 64 } \kappa_3  \Delta^{5/2}  (-x_0)^{-3/2}.
\label{o3dstr}
\end{eqnarray}
The gap $ \Delta (0)$ has by Eq.~(11) the strong-coupling
limit $ \Delta (0) \approx \epsilon_F [16/3 \pi ]^{3/2}|x_0|^{1/3}$
yielding the large-$x_0$ behavior

\begin{equation}
\frac{ \Omega}{V}  \sim  - \kappa_3 \epsilon_F^{5/2}
\frac{\pi}{64} \left(
\frac{16}{3\pi} \right)^{15/4}
 |x_0|^{-2/3}.
\end{equation}
In two dimensions, we substitute
 the gap
function
$\Delta$ of Eq.~(\ref{1.13}),
into the
 thermodynamic potential (\ref{o2t0}), and  obtain  for
 strong couplings an interesting result:
\begin{eqnarray}
\frac{ \Omega}{V}  \equiv 0
 \label{o2t0s}
\end{eqnarray}

Let us now turn to the entropy.
In three dimensions  near $T=0$ it is given for weak couplings  by:
\begin{eqnarray}
         { S \over V } =
         \kappa_3
        \sqrt{\mu}
\left\{
        \sqrt{ {  2 \pi \Delta ^3 \over T} }\exp
        \left(-
        {  \Delta \over T}
        \right)
\left[  1+ {\rm  erf}
        \left(
      \sqrt{  {\sqrt{ x_0^2+1}-1  \over T / \Delta   }}
\right)
\right]
 + 2 \mu
        \exp
        \left( -
        {  \sqrt{\mu^2 + \Delta^2} \over T }
        \right)
\right\},
\label{s}
\end{eqnarray}
For $ \mu / \Delta \rightarrow \infty$,
this reduces correctly  to the  BCS result:
\begin{eqnarray}
\frac{S}{V} =   \kappa_3  \sqrt{\mu}  \sqrt{ {  8 \pi \Delta ^3 \over T} }
\exp \left( -  { \Delta \over T} \right)
\label{sbcs}
\end{eqnarray}
In  two dimensions,
the  result is similar apart from the fact
that $ \Delta=\Delta(0)$ should be  given by Eq.~(\ref{1.13})
and $\kappa_3  \sqrt{\mu}$ should be replaced by
$\kappa_2$.
\comment{
\begin{eqnarray}
 { S \over V  } =   \kappa_2 \left\{ \sqrt{ {  2 \pi \Delta ^3 \over T} }
\exp \left( - { \Delta \over T} \right)
    \left[  1+ {\rm erf}  \left(
 \sqrt{  {\sqrt{ x_0^2+1}-1  \over T / \Delta   }}
 \right)\right]
+ 2 \mu
\exp \left( - { \sqrt{\mu^2 + \Delta^2} \over T }
   \right)  \right\},
\label{s2d}
\end{eqnarray}
The BCS limit of this is
\begin{eqnarray}
\frac{S}{V} =   \kappa_2 \sqrt{ {  8 \pi \Delta ^3 \over T} }
\exp \left( -  { \Delta \over T} \right)
\label{sbcs2}
\end{eqnarray}
}
In the strong-coupling limit
where  $ \mu/ \Delta \ll -1$,  we have for the entropy
in three dimensions
\begin{eqnarray}
\frac{S}{V}=  \kappa_3  { \sqrt{\pi} \over 4 }  T^{1/2 }
 \sqrt{\mu^2+\Delta^2}
  \exp\left( -
\frac{
  \sqrt{\mu^2+\Delta^2}}{T}
\right)
,
 \label{l1}
\end{eqnarray}
and in two dimensions:
\begin{eqnarray}
\frac{S}{V}=  - 2  \kappa_2 \mu
\exp\left( -\frac{  \sqrt{\mu^2+\Delta^2}}{ T}
\right)
\label{l2}
\end{eqnarray}
 From the entropy, we  easily derive
the
heat capacity at a constant volume
$c_V$.
In three dimensions it is given
near $T=0$  for weak and moderate
couplings by
\begin{eqnarray}
c_V=  \kappa_3 \sqrt{\mu}  \sqrt{2 \pi \Delta ^3} \left\{  { \Delta \over
T^{3/2} }
\exp \left( -{  \Delta \over T} \right) \left[ 1+ {\rm erf} \left(  \sqrt{
\frac{ \sqrt{x_0^2+1} -1  }{T/\Delta}} \right)
\right]
\right\}
\label{c3d}
\end{eqnarray}
reducing in the limit
 ${x_0 \rightarrow \infty }$
to the BCS result
\begin{eqnarray}
c_V =  \kappa_3 \sqrt{\mu}  \sqrt{2 \pi \Delta ^3} { 2\Delta \over T^{3/2} }
\exp \left(- {  \Delta \over T}  \right).
\label{c3dbcs}
\end{eqnarray}
In two dimensions, the weak-coupling behavior is
the same  with the factor
$\kappa_3  \sqrt{\mu}$  replaced by $\kappa_2$
\comment{
\begin{eqnarray}
c_V =  \kappa_2 \sqrt{2 \pi \Delta ^3} \left\{  { \Delta \over T^{3/2} }
\exp\left( - {  \Delta \over T} \right)  \left[ 1+
{\rm erf} \left(  \sqrt{
\frac{ \sqrt{x_0^2+1} -1  }{T/\Delta}} \right)
\right]
\right\},
\label{c2d}
\end{eqnarray}
}
while  the strong-coupling behavior in three dimensions is
\begin{eqnarray}
c_V =
\kappa_3  { \sqrt{\pi} \over 4 }  T^{-1/2 }
 (\mu^2+\Delta^2)
 \exp\left( -
\frac{
  \sqrt{\mu^2+\Delta^2}}{T}
\right)
\label{cstr3d}
\end{eqnarray}
and in two dimensions
\begin{eqnarray}
c_V = 2  \kappa_2   {\mu^2 \over T } \exp \left(
-\frac{\sqrt{\mu^2+\Delta^2}}{T}
\right).
\label{cstr2}
\end{eqnarray}
\section{ Conclusion}
The BCS-BEC crossover and related
phenomena is now a subject of
increasing interest in many branches of physics.
In this paper we have presented simple asymtotical results
for one of the important aspects of this phenomenon
in superconductors. 
Namely we studied
the behavior of thermodynamic functions when a system
crosses over from  the BCS superconductivity
to the Bose condensate of tightly bound pairs.
The first observation that could be made analyzing
the thermodynamic expressions is that 
the crossover
takes place in a quite a narrow region $1>x_0>-1$
which is in agreement with  previous studies.

While the crossover is known to
be continuous, the thermodynamic functions
 assume an exotic form in the regime
of strong attraction or low carrier density. The
interesting feature is
that a modified gap function $\sqrt{\Delta^2+\mu^2}$
enters all the expressions in the BEC limit in a similar way
as the gap function $\Delta$
enters thermodynamic expressions in the BCS limit.
Since in the BEC regime $\Delta <<-\mu$, 
the chemical potential is 
in some sense playing the same role
as the gap function in the BCS theory.
This result, that may seem surprising from the point of view
of the BCS theory, has the following roots:
The gap in the spectrum
of  single-particle excitations has a special feature \cite{Le}, \cite{Ea},
\cite{R-rev}
when the chemical potential changes its sign.
The sign change occurs at the minimum of the Bogoliubov
quasiparticle energy $E_{\bf k}$ where this energy
defines the gap energy in the quasiparticle spectrum:
\begin{eqnarray}
E_{\rm gap}={\rm min}\left(\xi_{\bf k}^2 +\Delta^2 \right)^{1/2} .
\label{1.5.1}
\end{eqnarray}
Thus, for positive chemical potential,
the gap energy is given directly by the gap function $ \Delta$, whereas for
negative chemical potential,
it is larger than that:
\begin{eqnarray}
E_{\rm gap}=  \Biggl\{ \matrix{
\Delta & {\rm for}  \ \ \ \mu >0 ,\cr
(\mu^2 +\Delta^2)^{1/2} & {\rm for} \ \ \ \mu < 0 .\cr}
\label{1.5.2}
\end{eqnarray}
That means that for the positive values of the 
chemical potential the gap function has the same meaning as 
the standard superconductive gap.
In contrast,  in the Bose limit,
when  chemical potential 
goes below the bottom of the band,
 one should distinguish between the
two notions of the ``gap". At first, there is  
the order parameter $\Delta$ that 
reaches zero at $T_c$ [ see e.g. \cite{R-rev}].
Also, in this limit, there exist the modified
gap function (\ref{1.5.2}) that tends to
$|\mu| = E_b/2$,  where $E_b=1/m a_s^2$ 
thus being proportional to the binding energy of the  pairs.
So when one is concerned 
about e.g. low- temperature thermodynamic functions
one should understand under the ``gap" basically  the chemical potential
rather than the order parameter $\Delta$.
Other well-known and to some extend similar features arising 
when a system crosses over from BCS to BEC limit are:
(i) the appearance of  two characteristic temperatures 
of different physical origin
$T_c$ and $T^*$ in BEC 
limit \cite{R8,Noz} in contrast
to only one characteristic temperature $T_c$ in 
the BCS limit; (ii)  
the  separation of notions of the Cooper pair
size and coherence length in strong coupling condensate
which in contrast coincide in the BCS limit
\cite{R8,pist,h}.

The above results for thermodynamical quantities once again 
illustrate the
 interesting circumstance of the essentially
richer nature
of a strong coupling condensate
comparing to a BCS condensate that helps to understand
better the phenomenon of symmetry breakdown in 
Fermi systems. 
\section{Acknowledgments}
The author is grateful to Prof. H. Kleinert for 
discussions and proofreading parts of the manuscripts 
and to Dr. D.M. Eagles for reporting a
typo in a previous version of the paper.
\renewcommand{\theequation}{\Alph{section}.\arabic{equation}}
\appendix
\section{Thermodynamics near critical temperature
of  superconductors with moderately strong coupling strength \label{App.A}}
In this Appendix we present expressions for the behavior
of the gap and the thermodynamic
functions near $T_c$ in superconductors
with moderately strong coupling strength. We employ a mean-field
approximation and thus our expressions are not valid
in the crossover region where one should incorporate
fluctuation corrections in the regime near $T_c$.

In the regime of  weak or moderately strong attraction and also
 for moderately low carrier density
the gap function behaves near $T_c$ as follows:
\begin{eqnarray}
 \left[
        { \Delta (T) \over  2T_c}
 \right]^2 =
{ \displaystyle
        \left( 1- { T \over T_c } \right)
 \left(
        1+ \tanh
        { \mu \over 2 T_c  }
 \right)
\over
    \displaystyle
        { 1 \over 4 }
 \left[ \displaystyle
          {1 \over \mu /2 T_c  } -
        {  \displaystyle
 1 \over (\mu /2 T_c )^2 }
        \tanh  { \mu \over 2 T_c  }
        \displaystyle
 \right] +
  \displaystyle
 \left(
        { 2 \over  \pi}
 \right) ^2 \
 \left(
        1 + \displaystyle
 { 2 \over \pi }
       \arctan
  \displaystyle
 \frac{\mu }{ \pi T_c}
 \right)
}.
\label{1tc1}
\end{eqnarray}
In the limit $ \lfrac {\mu }{ 2 T_c}  \rightarrow \infty$
this tends to the BCS result
\begin{eqnarray}
{\Delta(T) \over T_c} \simeq  3 \sqrt{1 - { T\over T_c} }.
\label{tc1bcs}
\end{eqnarray}
The ratio of zero temperature gap to the
critical temperature has the following dependence on
crossover parameter close to the BCS regime:
\begin{eqnarray}
{ \Delta (0) \over T_c } = { \pi \over e^\gamma }
\left(1-{\Delta (0)^2 \over 4 \mu^2} \right)^{-1/2}=
 { \pi \over e^\gamma }
\left(1-{1 \over 4 x_0^2 }\right)^{-1/2}
\label{n5}
\end{eqnarray}
Whereas the  critical temperature:
has the following dependence on $x_0$:
\begin{equation}
\frac{T_c}{\epsilon_F} \simeq
\frac{ e^\gamma}{ \pi  }
\left( \frac{1}{x_0} -{3 \over 8 x_0^3 }\right)
\end{equation}

The thermodynamic potential in three dimensions in the weak-coupling regime
 near $T_c $ is given
by
\begin{eqnarray}
\label{om3Dtc}
&& \!\!\!\!\!\!\!\!\!   \frac{ \Omega _s}{V} =
- \kappa_3 \sqrt{\mu }
\Biggl\{
        { (T_c -T) \Delta ^2 \over  2 T_c }
        \left[
        1+ \tanh
        {\mu \over 2 T_c}
        \right]+
 \\
&&     ~~~~     + {\Delta ^4 \over 4}
        { 1 \over (2 T_c )^2 }
\left[
        { 1 \over 4 }
        \left(
         {1 \over \mu /2T_c  }
        - { 1 \over (\mu /2 T_c )^2}
        \tanh
        {\mu \over 2 T_c}
        \right) +
        \left(
        { 2 \over  \pi}
         \right)  ^2
        \left(
        1 + { 2 \over \pi }
        \arctan
        \frac{\mu }{ \pi T_c}
        \right)
\right] \Biggr\}.
 \nonumber
\end{eqnarray}
In the  BCS limit, this reduces to the well-known formula:
\begin{eqnarray}
\frac{ \Omega _s}{V}
=  - \kappa_3 \sqrt{\mu}
\Delta ^2
\left(  1- { T \over  T_c } - { 1 \over 2 \pi ^2 }  { \Delta ^2  \over T_c ^2 }
 \right)
\label{ombc}
\end{eqnarray}
The entropy behaves near $T_c $ in three dimensions in the weak-coupling
regime like
\begin{eqnarray}
\frac{S_s}{V} \equiv  \frac{S-S_n}{V} =
- \kappa_3 \sqrt{\mu} {\Delta ^2   \over 2 T_c  }
\left[
1+ \tanh \left(  { \mu \over 2 T_c  } \right)
\right]
\label{s3dtc}
 \end{eqnarray}
with the BCS limit
\begin{eqnarray}
\frac{S_s}{V} =
-   \kappa_3 \sqrt{\mu} {\Delta ^2   \over  T_c  }.
\label{s3tcbcs}
\end{eqnarray}
In order
to derive the specific heat we must take into  account the
temperature dependence of the gap.

In three dimensions, we find in the
 weak-coupling region near $T_c$:
\begin{eqnarray}
 \frac{C_s}{V}  =
      2T  \kappa_3 \sqrt{ \mu}
{\displaystyle
 \left(
        1+ \tanh
        { \mu \over 2 T_c  }
 \right)^2
\over
    \displaystyle
  {  1 \over 4 }
 \left[
          {1 \over \mu /2T_c  } -
        { 1 \over (\mu /2 T_c )^2 }
        \tanh
        {\mu \over 2 T_c}
 \right] +
 \left(
        { 2 \over  \pi}
 \right) ^2 \
 \left(
        1 + { 2 \over \pi }
        \arctan
    \frac{\mu }{ \pi T_c}
\right)
},
\label{heattc3d}
\end{eqnarray}
which has the well-known BCS limit:
\begin{eqnarray}
 \frac{C_s}{V}  \simeq
  \kappa_3  \sqrt{ \mu} \pi^2 T_c .
\label{heatbcs}
\end{eqnarray}
 %
%

%

\end{document}